# Characterizing the Backscattered Spectrum of Mie Spheres

*Martín Molezuelas-Ferreras,\* Álvaro Nodar, María Barra-Burillo, Jorge Olmos-Trigo, Jon Lasa-Alonso, Iker Gómez-Viloria, Elena Posada, J. J. Miguel Varga, Rubén Esteban, Javier Aizpurua, Luis E. Hueso, Cefe Lopez, and Gabriel Molina-Terriza\**

This study describes both experimentally and theoretically an important hitherto undiscovered feature of the scattering of micron-sized spherical objects when illuminated with highly focused circularly polarized light. This is a regime of high experimental relevance which has not been described in full detail. The experiments are complemented with the analytical formulas explaining the field scattered directed toward the backward hemispace. In particular, it is proven that this field shows a very regular oscillatory dependency with the optical size. This phenomenon is typically hidden in the total scattered field, as the field is scattered much less toward the backward hemisphere than toward the forward one. These regular oscillations are measured experimentally. It is proven that, by analyzing them, it is possible to determine the index of refraction of isolated micron-sized particles, opening new paths for applications in sensing and metrology.

## 1. Introduction

The scattering of light from different objects is a fascinating field of research crucial to understanding effects in different areas of science, such as optics, astronomy, biology, and engineering, among many others. Interestingly, many scattering features can be deduced from the study of the interaction of light with highly symmetrical targets such as spherical objects. In 1908, Gustav Mie developed an analytical solution to the light scattering problem of a homogeneous sphere under plane-wave illumination.[1] This solution, widely known as Mie's theory,[2] is actively employed nowadays to understand the interaction between light and spherical particles.[3–5]

M. Molezuelas-Ferreras, Á. Nodar, J. Olmos-Trigo, J. Lasa-Alonso,
I. Gómez-Viloria, J. J. M. Varga, R. Esteban, J. Aizpurua, G. Molina-Terriza
Materials Physics Centre (CSIC-UPV/EHU)
Paseo Manuel de Lardizabal 5, Donostia / San Sebastián 20018, Spain
E-mail: mmolezuelas001@ikasle.ehu.eus; gabriel.molina@ehu.eus

M. Barra-Burillo, L. E. Hueso
CIC nanoGUNE
Tolosa Hiribidea 76, Donostia / San Sebastián 20018, Spain

J. Lasa-Alonso, J. J. M. Varga, R. Esteban, J. Aizpurua, C. Lopez, G. Molina-Terriza
Donostia International Physics Center (DIPC)
Paseo Manuel de Lardizabal 4, Donostia / San Sebastián 20018, Spain

J. Lasa-Alonso
Basic Sciences Department
Mondragon Unibertsitatea
Loramendi 4, Arrasate / Mondragón 20500, Spain

E. Posada, C. Lopez
Instituto de Ciencia de Materiales de Madrid
Sor Juana Inés de la Cruz 3, Madrid 28049, Spain

L. E. Hueso, G. Molina-Terriza
IKERBASQUE, Basque Foundation for Science
María Díaz de Haro 3, Bilbao 48013, Spain

The ORCID identification number(s) for the author(s) of this article can be found under https://doi.org/10.1002/lpor.202300665

DOI: 10.1002/lpor.202300665

Since then, numerous advancements and extensions have been made, including the so-called Generalized Lorentz-Mie theory (GLMT).[6–8] The GLMT provides a comprehensive approach for studying the optical response of spherical bodies at the nano- and micro-scale. An interesting application of this theory is the prediction of the selective excitation of individual multipolar modes.[9–11] By adequately adjusting intrinsic properties of the incident field, such as numerical aperture, total angular momentum,[12,13] and helicity of light,[14] the incident field can be precisely controlled. Therefore, the scattered field can be tailored by properly controlling the modes excited by a given incident field.[15–17] Moreover, the GLMT has driven multiple applications in the fields of particle characterization,[18] optical resonances,[19,20] harmonic generation,[21–23] and optical tweezers.[24–26]

In this research, we present both experimental and theoretical investigations into the phenomenon of backward scattering by spherical objects when subjected to highly focused beams. Within the scope of this study, we define backward scattering as the dispersion of electromagnetic fields directed toward the entirety of the scatterer's rearward hemisphere (i.e., to polar angles $\theta > 90°$). This approach extends beyond considering only fields propagating in the negative z-axis direction; instead, we encompass the entire semi-sphere. Similarly, we define forward scattering as the fields propagating into the forward hemisphere of the scatterer (i.e., $\theta < 90°$). The experimental data shows that the backward fields present a very regular oscillatory behavior. In the following section, we provide a full theoretical description of this phenomenon, including a closed-form analytical expression for the backward scattered power of dielectric spheres. Next, we present experimental measurements of the backscattered field, showing that it can be used to measure the refractive index of single isolated particles through the determination of the optical size.

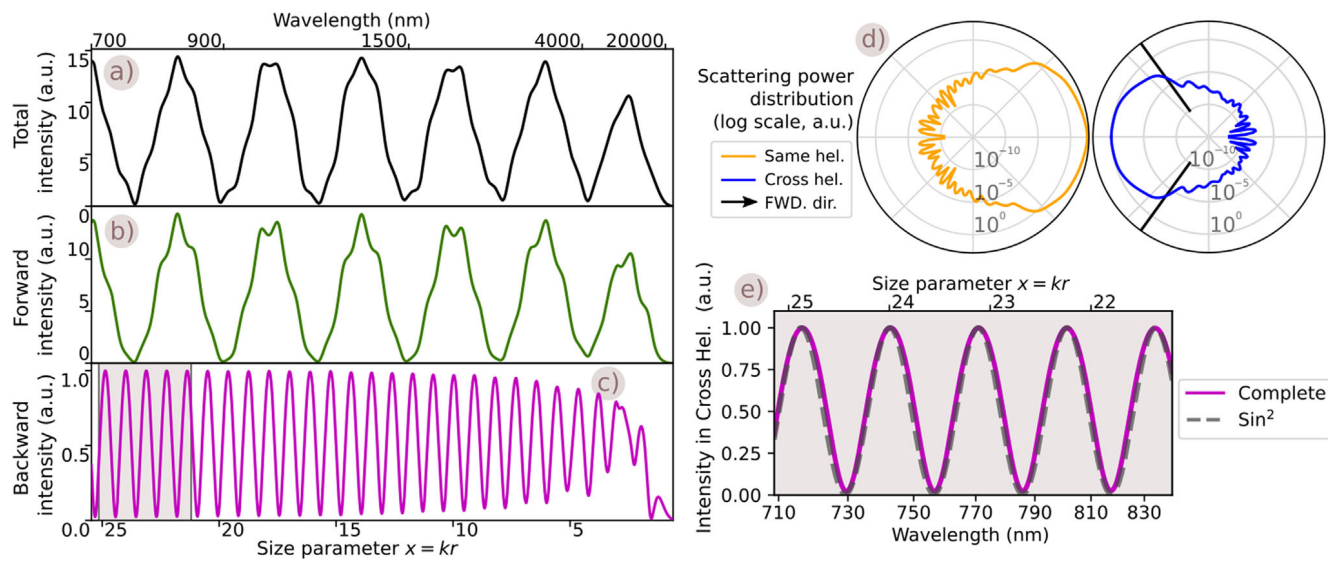

**Figure 1.** Analysis of the scattering of a Gaussian beam with positive helicity focused by a microscope objective with numerical aperture (NA) 0.81 on a dielectric sphere with refractive index 1.798 and radius 2.836μm, obtained numerically using GLMT (values are chosen this way to match those of the experiment). On a), the spectrum of total integrated intensity is calculated as a function of the size parameter $x = kr$. We also plot the numerically calculated intensity only in the forward (b) and backwards (c) hemi-spaces. On d), we present the numerically calculated scattering distribution in logarithmic scale of the scattering at $\lambda = 800$nm, (this wavelength is representative of the whole wavelength range studied experimentally) separated into the contribution with positive helicity (orange) and negative helicity (blue). The forward direction is to the right. On the cross helicity plot, a pair of radial black lines denote the NA of the microscope objective used on our experiment. On e), the shaded region highlighted in c) is presented with greater detail, for an easier comparison with experiments. The profile of the spectrum (purple) is fitted to a squared sine function (grey), using the analytical expression shown in the text.

## 2. Results and Discussion

### 2.1. Theoretical Description

As mentioned earlier, the GLMT provides a complete description of the overall scattering properties of particles. For example, it allows for finding analytical formulas for the total intensity of the scattered light off a particle. However, it is more challenging to find closed formulas which describe the emission of light onto a certain region. Generally speaking, finding the total power scattered in the forward or backward directions is a complex task except in some simple situations, like dipolar particles.[27–29] Here, we will show that in representative experimental situations, we can retrieve the backward scattered power and determine its overall properties.

**Figure 1** provides a numerical example of a typical experiment with a highly focused circularly polarized beam. In Figure 1a, we plot the total scattered power calculated for a dielectric Mie sphere versus the optical size parameter: $x = kr$, where $k = 2\pi/\lambda$ is the radiation wave-vector associated with the incident vacuum wavelength $\lambda$ and $r$ is the radius of the particle. The incident field is a circularly polarized Gaussian beam focused with a microscope objective with numerical aperture NA= 0.81. The scattering spectra are calculated using standard GLMT theory.[8,30] In Figure 1a, one can observe that the total scattered power presents an irregular pattern, seemingly composed by a sum of oscillations with different periods. On the other hand, in Figure 1b, we depict the scattered power by the same Mie sphere only in the forward direction. This is calculated by numerically integrating the intensity of the far-field propagating only in the forward direction. One can observe that the magnitude of the forward scattered power is very similar to the total scattered power. In this regard, comparing Figure 1a,b, we also notice that the scattered power in the forward direction shows the same kind of irregular oscillations present in the total scattered power with slight differences, but basically with the same period and phase.

In contrast, the scenario changes dramatically in Figure 1c, where we depict the backward scattered power by the same Mie sphere. Again, this is calculated by numerically integrating the far field, but this time in the backward directions. While its magnitude is much smaller than the forward and total scattered power counterparts, the backward scattered power presents regular oscillations with a more clearly defined frequency than the total and forward scattered power. Notice that the minima of the oscillations are close to zero backscattered power. Here, we provide a physical insight into the appearance of these oscillations, which, as we will see, can help us determine important properties of isolated particles.

A key element to understand this phenomenon is optical helicity,[31] which is defined as the projection of the angular momentum onto the linear momentum, $p = \mathbf{J} \cdot \mathbf{P}/P$, being $\mathbf{J}$ the angular momentum operator and $\mathbf{P}$ the linear momentum operator on electric fields. One way of understanding the helicity is in the plane wave decomposition of electric fields as it appears as the circular polarization of the plane waves. Therefore, only two helicity components of light are allowed, $p = +1$, in which all plane waves would have left circular polarization with respect to their propagation direction, and $p = -1$, where they would have right circular polarization. The two helicities of the scattered field propagate in a different manner. This can be observed in Figure 1d, where we plot in logarithmic scale separately the radiation pattern of the light scattered in the two helicities: scattered light with the same helicity as the incident field (left) and with the opposite helicity (right). Helicity on a paraxial beam[32] is directly equivalent to the right (−1) or left handedness (+1) of the beam's polarization, as it only contains a single propagation vector $k$. Thus, helicity can experimentally imprinted and filtered by manipulating the polarization of the beams. In a sort of spin-locking mechanism,[33] it can be observed that most of the same helicity radiation propagates in the forward direction, while the opposite helicity propagates preferentially backwards. This is a property of well-defined helicity multipolar modes[8,34] (see Supporting Information). Contrary to the typical spin-locking mechanism, this coupling of the propagation direction and helicity only appears in the scattered field. One way to observe this behavior is by using highly focused circularly polarized beams, composed only of lower order multipolar components.[8,11,31] Under these conditions, the backward (forward) intensity, can be approximated by only the opposite (same) helicity components ($P_{back} \approx P_{-p}$). Experimentally, we are limited by the numerical apperture of our focusing optics, but as we see also on Figure 1d, this parameter is large enough for us to retrieve the larger part of the scattering (> 99%) in our setup.

Thus, we proceed with our theoretical calculation by expanding a tightly-focused Gaussian beam with cylindrical symmetry onto multipolar modes of well-defined helicity. The symmetry of the problem determines the conservation of the $z$ projection of the angular momentum ($m_z$), both for the incident and outgoing modes. The incoming field also has a well-defined helicity ($p$), and the mode expansion is on the eigenstate of the total angular momentum, labeled $j$. While $m_z$ must be conserved in the scattering, the helicity is not generally preserved, giving rise to the same helicity ($p$) and opposite helicity ($-p$) components in the scattered field. Mie theory allows us to calculate the coefficients $\alpha_j = -(a_j + b_j)/2$ and $\beta_j = (a_j - b_j)/2$ representing, respectively the amplitude of helicity preservation and helicity flipping in the scattering process for each $j$ mode. Here, $a_j$, $b_j$ are the standard electric and magnetic Mie coefficients[2] (see Supporting Information). We will deal mainly with the backscattered field (the forward scattering can be retrieved similarly).

Therefore, as mentioned earlier, we consider the total power scattered in the opposite helicity of the incident field, $P_{-p}$, which depends only on the helicity coefficients $\beta_j$. This power is proportional to $P_{-p} \propto \sum_{j > m_z} \|C^p_{j,m_z}\|^2 \|\beta_j\|^2$. Here, $C^p_{j,m_z}$ are the amplitudes of the multipolar modes of the incident field. They depend on the shape and phase structure of the incident field and can be computed numerically (we present an example in **Figure 2**). However, as we will see shortly, for the case of highly focused fields and micron sized spheres, one does not need their exact values in order to analyze the periodicity of the scattering

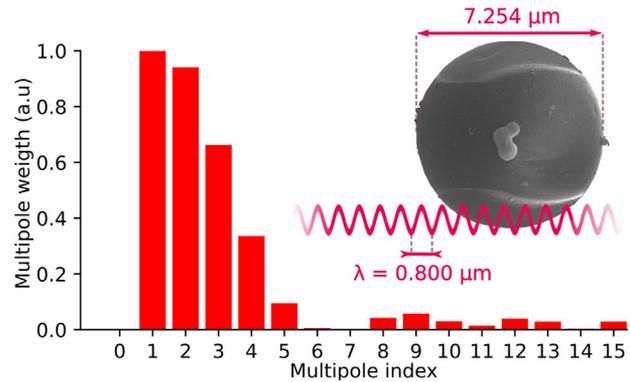

**Figure 2.** Multipolar decomposition of a highly focused Gaussian beam. On the graph is shown the multipole content of an input Gaussian beam with positive helicity, at $\lambda = 800$nm, focused by a microscope objective with NA= 0.81. The beam contains mainly a few multipoles of lower order (the small contribution of higher order harmonics is negligible compared to that of the lower order ones). Side by side, we also present one of the spheres used in the experiments and a scale of the wavelength.

spectrum. Let us now explicitly write down $\beta_j$, which can be expressed as:

$$\beta_j = \frac{(1 - m^2)\Psi'_j(mx)\Psi_j(mx)\left(\Psi'_j(x)\xi_j(x) - \Psi_j(x)\xi'_j(x)\right)}{2\left(m\Psi_j(mx)\xi'_j(x) - \xi_j(x)\Psi'_j(mx)\right)\left(\Psi_j(mx)\xi'_j(x) - m\xi_j(x)\Psi'_j(mx)\right)}$$

(1)

The $\Psi_j(z)$ and $\xi_j(z)$ functions are the standard Ricatti-Bessel functions of first and third order, respectively, which can be found in most standard photonic textbooks.[2] Here $x = kr$ is again the optical size parameter, and $m$ is the refractive index contrast. In the case that the sphere is surrounded by air $m \approx n$, with $n$ being the refractive index of the sphere. This is the case for our experimental conditions. We consider non-magnetic media, thus the magnetic permeability $\mu = 1$ for both the medium and the sphere. In general, to compute Equation (1), one needs to resort to numerical calculations due to its complexity. On the other hand, the expression can be simplified in the limit where $j << mkr$. Fortunately, this is a good approximation when considering a highly-focused beam impinging a microsphere. For tightly focused beams, the electromagnetic field can be approximated by a few lower order modes.[8] For example, in the case presented in Figure 2, although the contributions of multipolar modes with $j > 5$ to the focused beam are not strictly zero, their effect is very small and does not contribute appreciably to the incident field. In this regime, Equation (1) can be simplified using the asymptotic form of the Bessel functions. After some algebra (see Supporting Information), we arrive to an expression for the backscattered power:

$$P_{back} \approx P_{-p} \propto \sum_{j > m_z} \|C^+_{j,m_z}\|^2 \frac{(m^2 - 1)^2}{(m + 1)^4} \sin^2\left(2mx + \frac{j(j + 1)}{mx}\right)$$

(2)

Notice that this expression can be further simplified if one can neglect the $j(j + 1)/mx$ phase inside the squared sin function,

2300665 (3 of 7)

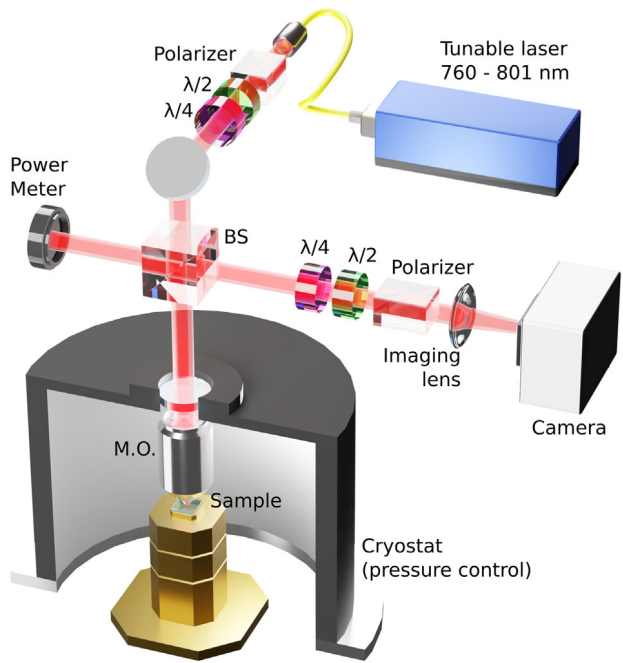

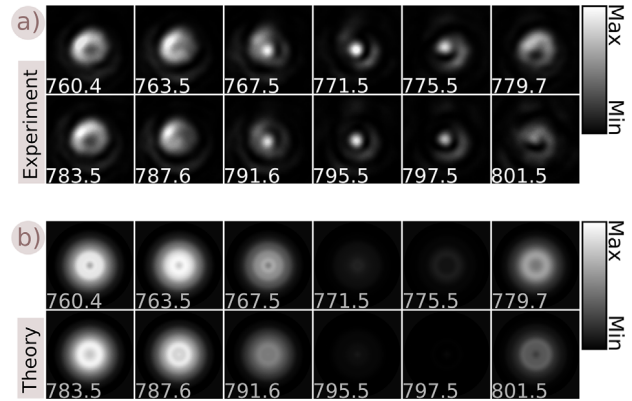

**Figure 4.** On a), captured images of the cross-helicity component of the beam scattered by one of the spheres (depicted in Figure 5a), at different wavelengths over the spectrum. On b), the calculated intensity profiles for the same sphere, using Mie Theory, for the same experimental conditions. Number labels in white and grey stand for wavelengths in nm. The scale of the plots is linear, and the maximum is chosen as the greatest intensity value on each set of images. Each image is normalized in this manner.

**Figure 3.** Experimental setup. We use a fiber-coupled tunable diode laser of wavelength 760–801 nm and control its polarization by a set of a polarizer, a quarter wave plate (QWP), and a half wave plate (HWP). Together they are able to create any polarization on the back focal plane of the microscope objective (MO). A similar but reversed arrangement is used at the output, but this time to filter alternatively the polarization components of scattered light. The beam is focused on the sample using a MO, which also collects the scattered light. The input power is controlled by a power meter. The sample and the MO are contained within the chamber of a cryostat, which provides pressure control. The pressure was kept at $10^{-4}$ mbar, and the temperature at 394K (ambient). The input and output beams are separated by a beam splitter (BS), and the output scattered beam is imaged by another lens on a CCD camera after the polarization filtering.

which is the typical case in our range of measurements. In this case, the backscattered power takes the simple form of $P_{back} \propto \sin^2(2mx + \varphi)$, where $\varphi$ accounts for a small overall phase. Equation (2) is the main theoretical result of the present work. It describes the backscattered power of a Mie sphere in terms of a sinusoidal function with regular oscillations dependent on its optical size, as we previously anticipated. Now, by a proper fitting of the frequency of the sin function (see Figure 1e), we can obtain intrinsic properties of the Mie sphere, such as the optical size or the refractive index contrast.

## 2.2. Experiments and Analysis

We turn now to the experimental measurements of the backscattered power from several dielectric microspheres with highly focused beams with a well-defined helicity. We use the previously developed theoretical analysis in order to accurately fit the experimental data. The setup used for this task is depicted in **Figure 3**. A tunable laser (760–801nm) was focused onto micrometric $TiO_2$ spheres on a glass substrate using a microscope objective (MO), which also collects the scattered light. The scattered light was then imaged onto a CCD camera recording the intensity profile of the field. The backscattered power was obtained by summing over the pixels of each frame. The polarization of the input was controlled at the back focal plane of the MO. Both the input and the measured scattered fields have right-handed circular polarization, but due to their different propagation directions, they had opposite helicity (for more details about the experimental setup and measurements, see Supporting Information).

After registering the scattering at the setup, we measured the diameter of the spheres to a precision of tens of nanometers using Scanning Electron Microscopy (SEM) (Figure 5). On the other hand, as it was previously noted, the oscillatory pattern of the backscattered power with respect to the wavelength allows for obtaining the optical size of the particle. Therefore, this procedure can be used to measure the effective refractive index of each single sphere.

In **Figure 4**a, we present the experimental images of the spatial profile of the back-scattered field for the sphere in **Figure 5**a. While the recorded scattering profiles are not fully cylindrically symmetric, due to slight defects on the spheres, one can notice that the intensity and the shape of the field change noticeably over the range of wavelengths on a nearly cylindrical manner. Also, in Figure 4b, the calculated scattered field profile from GLMT is shown for a sphere of this size with the beam multipolar mode expansion calculated in Figure 2. One can observe that the Mie scattering model captures the features not only of the overall intensity, but also of the spatial profile. The main differing detail between the experimental and calculated intensity profiles is a constant Gaussian spot on the middle of every experimental measurement, that can be attributed to the reflection of the beam on the sample substrate. Note that on the numerically calculated plots labeled with 771.5, 775.5, 795.5, and 797.5, the profile is not fully dark, but rather very dim in comparison with the brighter plots, and also in contrast with the correspondent experimental plots that show a Gaussian spot. While the reflection from the substrate is not considered in the theory, it can be easily subtracted and normalized in the experimental measurements (see Supporting Information).

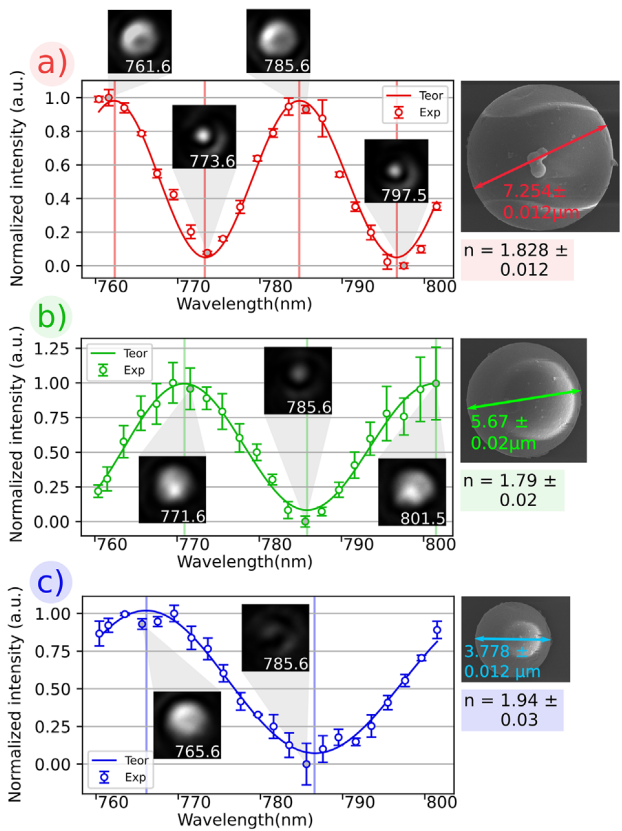

**Figure 5.** a-c) Scattering spectra for the TiO$_2$ spheres shown on the right. The points account for the experimental data, and the solid lines for the fit of these data to the squared sine function predicted on Equation (2). The minimum and maximum values of the spectra are rescaled so that the minimum intensity equates to zero and the maximum intensity equates to one. Each curve is accompanied by the SEM image of the corresponding sphere and its measured diameter. The value of the adjusted refractive index (n) is also shown. The spatial profile of the backscattered field intensity is displayed for the points that are closer to the crests and valleys of each oscillation. On the field images, the number label in white stands for the wavelength in nm at which it was taken. As in **Figure 4**, the scale of each of the field images is linear, and the maximum is chosen as the greatest intensity value on each set of images.

Now, we can represent the total backscattered power with respect to the wavelength for every sphere, and compare it with the theoretically expected oscillatory behavior (Figure 5). In this figure, we also provide a correspondence of the intensity minima and maxima with the actual image of the scattered field at these wavelengths. On the points of maximum intensity, the fields are much more spread and intense than in the lowest intensity points, in which the contribution is darker and mainly dominated by the reflection on the substrate.

In Figure 5, the experimental results are fitted to the expected simple oscillatory pattern as derived in Equation (2). The measured value of the radius of the spheres allows us to obtain a value for the refractive index of each individual particle. The uncertainty of these measurements was determined by the number of complete cycles that one could observe on a given spectral range and the experimental error of the measurements. The solid line in Figure 5 corresponds to the best-fitted index of refraction. It can be observed that the fitted curves and the experimental points match nicely. The values of the effective refractive index of each particle adjust to what could be expected from large amorphous TiO$_2$ spheres,[35] and the disparity of values is in agreement with the variation that can occur in the different processes of growth during the same synthesis process for each of the spheres.

The determination of the refractive index of spherical particles is a complicated experimental problem that has been approached in different ways in the literature, all presenting certain restrictions. If one can average over many spheres, some of the methods that have been used are the analysis of the diffuse reflectance and transmittance of particle suspensions,[36] the refractive index matching of the media surrounding the particles,[37] the characterization of the light reflected off the Bragg planes of spherical particle clusters,[38,39] or holographic microscopy applied to a polydisperse nanoparticle suspension.[40] Nonetheless, these methods cannot be used for the characterization of a single isolated particle. This can be done by using holographic techniques,[41] or by means of an optical trapping setup by fitting the dynamic behavior of a spherical particle[42] or a biological sample.[43] A review of these and other methods can be found in ref. [44]. Our work adds an innovative way of determining the refractive index and size of single micrometric particles with great precision, on a regime of sizes and refractive indices that is not easily accessed by any of the previous methods and where the particle could lie on a substrate so that it can be easily retrieved.

Moreover, the simplicity of the experimental setup allows for an easy implementation on other spherical structures. A wider wavelength range, or the utilization of Laguerre-Gaussian beams with a different multipole content,[8] are some further upgrades that can be added to a similar experimental setup in order to refine the measurements and make them viable for a wider range of particle sizes and materials. The role of the sample substrate on the results of the study is also something to add to a further theoretical modeling of this experiment. Also, as the sample environment is not altered by the measurement and could be very flexible, this method also allows for the continuous monitoring of the refractive index of spherical objects in changing environments, such as chemical synthesis or pressure and temperature changes.

## 3. Conclusion

In conclusion, we have derived a closed-analytical expression for the backscattered power of multipolar Mie spheres when illuminated with a tightly-focused Gaussian beam with well-defined helicity. We have been able to reduce the obtained expression to an oscillatory pattern, in which the period and the optical size of the particles are tied. We have also measured this oscillatory pattern on several dielectric spheres of known size, and upon analysis we have been able to extract the value of their refractive index. The agreement between theory and experiment is remarkable, underscoring the validity of our approach.

## Supporting Information

Supporting Information is available from the Wiley Online Library or from the author.

## Acknowledgements

M.M.F acknowledges financial support from MCIN/AEI/10.13039/501100011033 and "ESF Investing in your Future" through Project Ref. No PRE2018-085136. M.M.F. and G.M.T. acknowledge financial support from MCIN/AEI/10.13039/501100011033 through Project Ref. No. FIS2017-87363-P. Á.N. acknowledges financial support from MCIN/AEI/10.13039/501100011033 and "ESF Investing in your Future" through Project Ref. No. BES-2017-080073. Á.N., R.E., and J.A. acknowledge financial support from MCIN/AEI/10.13039/501100011033 and "ERDF A way of making Europe" through Project Ref. No. PID2022-139579NB-L00. M.M.F., Á.N., J.J.M.V., R.E., J.A. and G.M.T. acknowledge financial suport from the Department of Education, Research and Universities of the Basque Government through Project Ref. No. IT 1526-22. M.M.F., J.L.A., I.G.V., J.J.M.V. and G.M.T. acknowledge financial support from CSIC Research Platform PTI-001. M.B.B. acknowledges financial support from MCIN/AEI/10.13039/501100011033 through Project Ref. No. MDM-2016-0618. J.O.T. acknowledges financial support from MCIN/AEI/10.13039/501100011033 and the European Union NextGenerationEU/PRTR through the Juan de la Cierva Fellowship Ref. No. FJC2021-047090-I, and from MCIN/AEI/10.13039/501100011033 and "ERDF A way of making Europe" through Project Ref. No. PID-2022-137569NB-C43. C.L. acknowledges financial support from MCIN/AEI/10.13039/501100011033 through Project Ref. No. PID2021-124814NB-C21.

## Conflict of Interest

The authors declare no conflict of interest.

## Author Contributions

M.M.F., Á.N., R.E., J.A. and G.M.T. conceived the original idea. G.M.T. derived the expressions for the scattered power from GLMT. M.M.F, Á.N., and G.M.T. performed the computer simulations of the theoretical models. M.M.F., J.J.M.V., I.G.V., and J.L.A. built the experimental setup. M.M.F carried out the scattering measurements and analyzed the data. C.L and E.P. fabricated the spherical particles. M.B.B and L.H. performed the SEM measurements. M.M.F, Á.N., J.O.T., C.L., and G.M.T. contributed to the interpretation of the results. M.M.F, Á.N., J.O.T., and G.M.T. wrote the article with the input from the other authors.

## Data Availability Statement

The data that support the findings of this study are available from the corresponding author upon reasonable request. The data used for the figures and results in this article can be found at the digital.csic.es repository.